\documentclass[12pt,thmsa]{article}
\usepackage{amssymb}

\usepackage{sw20aip}

\input{tcilatex}
\begin{document}

\title{INHOMOGENEOUS MAGNETIC RESPONSE IN ANYON FLUID AT HIGH TEMPERATURE}
\author{\textit{E. J. Ferrer and V. de la Incera} \and \textit{Dept. of Physics,
State University of New York, Fredonia, NY 14063, USA}}
\date{SUNY-FRE-98-10}
\maketitle

\begin{abstract}
The charged anyon fluid in the presence of an externally applied constant
and homogeneous magnetic field is investigated at temperatures larger than
the energy gap ($T\gg \omega _{c}$). It is shown that the applied magnetic
field inhomogeneously penetrates the sample with a spatial periodicity
depending on a wavelength that decreases with temperature. The distribution
of charges in the ($T\gg \omega _{c}$)-phase acquires a periodic spatial
arrangement.
\end{abstract}

Although currently there is no direct experimental evidence of the existence
of anyons, it has been argued that strongly correlated electron systems in
two dimensions can be associated with the existence of anyonic
quasi-particles\cite{5}$.$ Anyons \cite{Anyons}$^{,}$ \cite{Wilczek} are
particles with fractional statistics in (2+1)-dimensions. The anyon
description within the Chern-Simons (CS) gauge theory is equivalent to
attaching flux tubes to the charged fermions. The fractional exchange
statistics \cite{Wilczek} that characterizes anyons comes from the
Aharonov-Bohm phases resulting from the adiabatic transport of two anyons$.$

Charged anyons exhibit the remarkable property of superconductivity \cite
{Sup}$^{,}$\cite{Chen}$.$ Anyon superconductivity's origin is different from
the Nambu-Goldstone-Higgs like mechanism\cite{Wilczek}$.$ The genesis of the
anyon superconductivity is given by the spontaneously violation of
commutativity of translations in the free anyon system\cite{Chen}$.$ This
new mechanism might find wide applications in new physical situations and
deserves a deeper analysis.

Up to recent years the charged anyon fluid was considered to superconduct
only at $T=0$ \cite{Sup}$^{-}$\cite{Arova}$,$ because the Meissner effect
seemed to disappear at any finite temperature \cite{9}$^{-}$\cite{11}$.$
However, as shown in previous papers \cite{Our}$,$ boundary effects can
affect the dynamics of this two-dimensional system in such a way that the
long-range mode \cite{8}$,$ which accounts for a homogeneous magnetic field
penetration \cite{11}$,$ and thus for the lack of a Meissner effect, is
forbidden in the bounded sample.

In the present work we show that at temperatures larger than the energy gap (%
$T\gg \omega _{c}$) a non-superconducting phase exists in the charged anyon
fluid. This result, indicates the existence of a phase transition in this
fluid from a superconducting phase, at $T\ll \omega _{c}$ (see Ref. [10]),
to a non-superconducting phase, at $T\gg \omega _{c}$.

In the temperature ranges corresponding to these two phases there is no
mathematical singularity in physical quantities, as other authors have also
pointed out \cite{8}$^{,}$\cite{11}$.$ However, taking into account that the
order parameter that separates these two phases is the energy gap $\omega
_{c}$, it is logic to expect that any oddity in the behavior of the physical
quantities (like a divergence in the penetration length), which would count
for a phase transition, should not manifest itself in temperature regions
far away from the order of magnitude of the energy gap (i.e. at $T\ll \omega
_{c}$, or $T\gg \omega _{c}$). In order to observe a signature of the phase
transition in the physical quantities we would need to consider a different
temperature approximation, i.e. $T\sim \omega _{c}$.

The evaporation of the superconducting state at $T\gg \omega _{c}$ is
however a natural result. At those temperatures the electron thermal
fluctuations should make accessible the free states existing beyond the
energy gap. As a consequence, the charged anyon fluid should not be a
perfect conductor at those temperatures. An indication of such a transition
may be found studying the system magnetic response at $T\gg \omega _{c}$.
Our main goal in this paper is to probe this region by investigating the
characteristics of the magnetic response of the system to an applied
constant and homogeneous magnetic field.

Below, we show that at $T\gg \omega _{c}$ that externally applied magnetic
field can penetrate the sample, giving rise to a periodically inhomogeneous
magnetic field within the bulk. The inhomogeneity of the magnetic response
increases with the temperature. Moreover, we prove that the inhomogeneous
character of the magnetic response in the high temperature phase is linked
to the inhomogeneity of the induced many-particle charge and current
densities at $T\gg \omega _{c}$.

The periodic inhomogeneity of the charge density, which takes place in the
new phase, may be associated with a sort of Wigner transition\cite{T}$.$ As
Wigner\cite{Wigner} pointed out, a system of interacting electrons may form
a lattice when their density is lowered (liquid-crystal transition). In the
charged anyon system, when the temperature is larger than the energy gap,
the electron density per Landau level is decreased. This decrease makes
energetically more favorable the periodic structure (crystallization) of the
system.

Let us study the linear magnetic response of the charged anyon fluid to an
applied constant and uniform magnetic field at temperatures larger than the
energy gap. The linear response of the medium can be found under the
assumption that the quantum fluctuations of the gauge fields about the
ground-state are small. In this case the one-loop fermion contribution to
the effective action, obtained after integrating out the fermion fields, can
be evaluated up to second order in the gauge fields. The effective action in
terms of the quantum fluctuations of the gauge fields within the linear
approximation takes the form\cite{8}$^{-}$\cite{Our}

\begin{equation}
\Gamma _{eff}\,\left( A_{\nu },a_{\nu }\right) =\int dx\left( -\frac{1}{4}%
F_{\mu \nu }^{2}-\frac{N}{4\pi }\varepsilon ^{\mu \nu \rho }a_{\mu }\partial
_{\nu }a_{\rho }\right) +\Gamma ^{\left( 2\right) }  \tag{1}
\end{equation}
$\Gamma ^{\left( 2\right) }$ is the one-loop fermion contribution to the
effective action in the linear approximation

\begin{equation}
\Gamma ^{\left( 2\right) }=\int dxdy\left[ a_{\mu }\left( x\right) +eA_{\mu
}\left( x\right) \right] \Pi ^{\mu \nu }\left( x,y\right) \left[ a_{\nu
}\left( y\right) +eA_{\nu }\left( y\right) \right] .  \tag{2}
\end{equation}

In Eq. (2) $\Pi _{\mu \nu }$ represents the fermion one-loop polarization
operator in the presence of the CS background magnetic field $\overline{b}$,
which appears, as it is known \cite{8}$^{-}$\cite{Our}$,$ to guarantee the
system neutrality in the presence of an electron finite density.

As we are interested in the response of the system to a uniform and constant
applied magnetic field, it is enough to consider the leading behavior of $%
\Pi _{\mu \nu }$ for static $\left( k_{0}=0\right) $ and slowly $\left( 
\mathbf{k}\sim 0\right) $ varying configurations. In this limit, using the
frame on which $k^{i}=\left( k,0\right) $, $i=1,2$, the polarization
operator takes the form

\begin{equation}
\Pi ^{\mu \nu }=\left( 
\begin{array}{ccc}
-\left( \mathit{\Pi }_{\mathit{0}}+\mathit{\Pi }_{\mathit{0}}\,^{\prime
}\,k^{2}\right) & 0 & -i\mathit{\Pi }_{\mathit{1}}k \\ 
0 & 0 & 0 \\ 
i\mathit{\Pi }_{\mathit{1}}k & 0 & \mathit{\Pi }_{\,\mathit{2}}k^{2}
\end{array}
\right)  \tag{3}
\end{equation}

The leading contributions of the one-loop polarization operator coefficients 
$\mathit{\Pi }_{\mathit{0}}$, $\mathit{\Pi }_{\mathit{0}}\,^{\prime }$, $%
\mathit{\Pi }_{\mathit{1}}$ and $\mathit{\Pi }_{\,\mathit{2}}$ in the static
limit ($k_{0}=0,$ $\mathbf{k}\sim 0$) at high temperatures $\left( T\gg
\omega _{c}\right) $ are

\[
\mathit{\Pi }_{\mathit{0}}=\frac{m}{2\pi }\left[ \tanh \frac{\beta \mu }{2}%
+1\right] ,\qquad \mathit{\Pi }_{\mathit{0}}\,^{\prime }=-\frac{\beta }{%
48\pi }\func{sech}\!^{2}\!\,\left( \frac{\beta \mu }{2}\right) ,\qquad 
\]

\begin{equation}
\mathit{\Pi }_{\mathit{1}}=\frac{\overline{b}}{m}\mathit{\Pi }_{\mathit{0}%
}\,^{\prime },\qquad \mathit{\Pi }_{\,\mathit{2}}=\frac{1}{12m^{2}}\mathit{%
\Pi }_{\mathit{0}}  \tag{4}
\end{equation}
In these expressions $\mu $ is the chemical potential and $m=2m_{e}$ ($m_{e}$
is the electron mass). These results are in agreement with those found in
Refs.[8] and [13].

The extremum equations obtained from the effective action (1) for the
Maxwell and CS fields are

\begin{equation}
\mathbf{\nabla }\cdot \mathbf{E}=eJ_{0}  \tag{5}
\end{equation}

\begin{equation}
-\partial _{0}E_{k}+\varepsilon ^{kl}\partial _{l}B=eJ^{k}  \tag{6}
\end{equation}

\begin{equation}
\frac{eN}{2\pi }b=\mathbf{\nabla }\cdot \mathbf{E}  \tag{7}
\end{equation}

\begin{equation}
\frac{eN}{2\pi }f_{0k}=\varepsilon ^{kl}\partial _{0}E_{l}+\partial _{k}B 
\tag{8}
\end{equation}
$f_{\mu \nu }$ is the CS gauge field strength tensor, defined as $f_{\mu \nu
}=\partial _{\mu }a_{\nu }-\partial _{\nu }a_{\mu }$, and $J_{ind}^{\mu }$
is the current density induced by the many-particle system.

\begin{equation}
J_{ind}^{0}\left( x\right) =\mathit{\Pi }_{\mathit{0}}\left[ a_{0}\left(
x\right) +eA_{0}\left( x\right) \right] +\mathit{\Pi }_{\mathit{0}%
}\,^{\prime }\partial _{x}\left( \mathcal{E}+eE\right) +\mathit{\Pi }_{%
\mathit{1}}\left( b+eB\right)  \tag{9}
\end{equation}

\begin{equation}
J_{ind}^{1}\left( x\right) =0,\qquad J_{ind}^{2}\left( x\right) =\mathit{\Pi 
}_{\mathit{1}}\left( \mathcal{E}+eE\right) +\mathit{\Pi }_{\,\mathit{2}%
}\partial _{x}\left( b+eB\right)  \tag{10}
\end{equation}
In the above expressions we used the following notation: $\mathcal{E}=f_{01}$%
, $E=F_{01}$, $b=f_{12}$ and $B=F_{12}$. We confine our analysis to gauge
field configurations which are static and uniform in the y-direction. Within
this restriction we are taking a gauge in which $A_{1}=a_{1}=0$.

The magnetic and electric field solutions obtained from Eqs. (5)-(8) are
respectively

\begin{equation}
B\left( x\right) =-\gamma _{1}\left( C_{1}e^{-x\xi _{1}}-C_{2}e^{x\xi
_{1}}\right) -\gamma _{2}\left( C_{3}e^{-x\xi _{2}}-C_{4}e^{x\xi
_{2}}\right) +C_{5}  \tag{11}
\end{equation}
\begin{equation}
E\left( x\right) =C_{1}e^{-x\xi _{1}}+C_{2}e^{x\xi _{1}}+C_{3}e^{-x\xi
_{2}}+C_{4}e^{x\xi _{2}},  \tag{12}
\end{equation}
where $\gamma _{1}=\left( \xi _{1}^{2}\kappa +\eta \right) /\xi _{1}$, $%
\gamma _{2}=\left( \xi _{2}^{2}\kappa +\eta \right) /\xi _{2}$, and the
inverse length scales $\xi _{1}$ and $\xi _{2}$ are

\[
\xi _{1,2}=\left[ -d\pm \sqrt{d^{2}-4ac}\right] ^{\frac{1}{2}}/\sqrt{2a} 
\]

\begin{equation}
a=\omega \kappa ,\qquad c=\alpha \eta -\sigma \gamma -\tau \eta \chi ,\qquad
d=\omega \eta +\alpha \kappa -\gamma -\tau \kappa \chi  \tag{13}
\end{equation}
In these equations the temperature enters through the polarization operator
coefficients as

\[
\omega =\frac{2\pi }{N}\mathit{\Pi }_{\mathit{0}}\,^{\prime },\quad \alpha
=-e^{2}\mathit{\Pi }_{\mathit{1}},\quad \tau =e\mathit{\Pi }_{\mathit{0}%
},\quad \chi =\frac{2\pi }{eN},\quad \sigma =-\frac{e^{2}}{\gamma }\mathit{%
\Pi }_{\mathit{0}},\quad \eta =-\frac{e^{2}}{\delta }\mathit{\Pi }_{\mathit{1%
}}, 
\]

\begin{equation}
\gamma =1+e^{2}\mathit{\Pi }_{\mathit{0}}\,^{\prime }-\frac{2\pi }{N}\mathit{%
\Pi }_{\mathit{1}},\quad \delta =1+e^{2}\mathit{\Pi }_{\,\mathit{2}}-\frac{%
2\pi }{N}\mathit{\Pi }_{\mathit{1}},\quad \kappa =\frac{2\pi }{N\delta }%
\mathit{\Pi }_{\,\mathit{2}}.  \tag{14}
\end{equation}

As it is known, for the determination of the magnetic response (11) it is
essential to investigate the nature of the inverse length scales at the
temperatures of interest. For example, in the $T\ll \omega _{c}$ phase, the
real character of the inverse length scales $\xi _{1}$ and $\xi _{2}$ was
crucial for the realization of the Meissner effect (see Ref. [10]). At
temperatures much larger than the energy gap $\left( T\gg \omega _{c}\right) 
$ the leading contribution to the inverse length scales are given by

\begin{equation}
\xi _{1}\simeq e\sqrt{\mathit{\Pi }_{\mathit{0}}}=e\sqrt{m/2\pi }\left(
\tanh \frac{\beta \mu }{2}+1\right) ^{\frac{1}{2}}  \tag{15}
\end{equation}

\begin{equation}
\xi _{2}\simeq \frac{1}{\pi }\left( \mathit{\Pi }_{\,\mathit{2}}\mathit{\Pi }%
_{\mathit{0}}\,^{\prime }\right) ^{-1/2}=24i\sqrt{2m/\beta }\cosh \frac{%
\beta \mu }{2}\left( \tanh \frac{\beta \mu }{2}+1\right) ^{-\frac{1}{2}} 
\tag{16}
\end{equation}

From Eqs. (15) and (16) we have that $\xi _{1}$ is real, but $\xi _{2}$ is
imaginary. The imaginary value of the inverse length $\xi _{2}$ is due to
the fact that at $T\gg \omega _{c}$, $\mathit{\Pi }_{\,\mathit{2}}>0$ and $%
\mathit{\Pi }_{\mathit{0}}\,^{\prime }<0$ (see Eq. (4)). An imaginary $\xi
_{2}$ implies that the term $\gamma _{2}\left( C_{3}e^{-x\xi
_{2}}-C_{4}e^{x\xi _{2}}\right) $, in the magnetic field solution (11), does
not have a damping behavior, but an oscillating one.

The magnetic response is completely determined only after we find the values
of the unknown coefficients $C^{\prime }s$ appearing in Eq. (11). In doing
that, we must take into account the problem boundary conditions and the
minimization of the system free-energy density. We should point out that the
presence of the constant coefficient $C_{5}$ in Eq. (11) is associated with
the possible propagation of a magnetic long-range mode.

Henceforth we consider the anyon fluid confined to a half plane $-\infty
<y<\infty $ with boundary at $x=0$. The external magnetic field is applied
from the vacuum ($-\infty <x<0$). The boundary conditions for the magnetic
field are $B\left( x=0\right) =\overline{B}$ ($\overline{B}$ constant), and $%
B\left( x\rightarrow \infty \right) $ finite. Since in Eqs. (5)-(8) the
magnetic field is tangled to the electric field $E$, the boundary values of $%
E$ have to be taken into account in determining the unknown coefficients.
Because no external electric field is applied, the boundary conditions for
this field are, $E\left( x=0\right) =0$, $E\left( x\rightarrow \infty
\right) $ finite. After using these conditions for $B$ and $E$ it is found
that,

\begin{equation}
C_{2}=0,\qquad -C_{1}=C_{3}+C_{4},\qquad C_{1}=\frac{C_{5}-(C_{3}-C_{4})%
\gamma _{2}-\overline{B}}{\gamma _{1}}  \tag{17}
\end{equation}

Using Eqs. (17) the magnetic and electric fields (Eqs. (11) and (12)
respectively) can be written in the following convenient form

\begin{equation}
B\left( x\right) =-\gamma _{1}C_{1}e^{-x\xi _{1}}+\overline{\gamma }%
_{2}\left( A\cos x\overline{\xi }_{2}+C_{1}\sin x\overline{\xi }_{2}\right)
+C_{5}\text{ }  \tag{18}
\end{equation}

\begin{equation}
E\left( x\right) =C_{1}e^{-x\xi _{1}}-C_{1}\cos x\overline{\xi }_{2}+A\sin x%
\overline{\xi }_{2}  \tag{19}
\end{equation}
where

\begin{equation}
A=i\left( C_{4}-C_{3}\right) ,\qquad \overline{\gamma }_{2}=\left( \overline{%
\xi }_{2}{}^{2}\kappa -\eta \right) /\overline{\xi }_{2},\qquad \xi _{2}=i%
\overline{\xi }_{2}  \tag{20}
\end{equation}

From Eqs. (18) and (19) we see that to determine $C_{1}$, $C_{5}$ and $A$
one needs to consider another physical constraint besides the boundary
conditions (17). Since, obviously, any meaningful solution have to be
stable, a natural additional condition is given by the stability equation
derived from the system free energy. To obtain it, we start from the free
energy of the half-plane sample

\[
\mathcal{F}=\frac{1}{2}\int\limits_{-L^{\prime }/2}^{L^{\prime
}/2}dy\int\limits_{0}^{L}dx\left\{ \left( E^{2}+B^{2}\right) +\frac{N}{\pi }%
a_{0}b-\mathit{\Pi }_{\mathit{0}}\left( eA_{0}+a_{0}\right) ^{2}\right. 
\]

\begin{equation}
\left. -\mathit{\Pi }_{\mathit{0}}\,^{\prime }\left( eE+\mathcal{E}\right)
^{2}-2\mathit{\Pi }_{\mathit{1}}\left( eA_{0}+a_{0}\right) \left(
eB+b\right) +\mathit{\Pi }_{\,\mathit{2}}\left( eB+b\right) ^{2}\right\} 
\tag{21}
\end{equation}
where $L$ and $L^{\prime }$ determine the two sample's lengths.

In Eq. (21) we have to substitute the field solutions (18) and (19) together
with the solutions for the CS fields (that can be found substituting Eqs.
(18) and (19) in Eqs. (7) and (8))

\begin{equation}
b\left( x\right) =\chi \xi _{1}C_{1}e^{-x\xi _{1}}-\chi \overline{\xi }%
_{2}\left( A\cos x\overline{\xi }_{2}+C_{1}\sin x\overline{\xi }_{2}\right) 
\tag{22}
\end{equation}

\begin{equation}
\mathcal{E}\left( x\right) =-\chi \xi _{1}\gamma _{1}C_{1}e^{-x\xi
_{1}}+\chi \overline{\xi }_{2}\overline{\gamma }_{2}\left( A\sin x\overline{%
\xi }_{2}-C_{1}\cos x\overline{\xi }_{2}\right) .  \tag{23}
\end{equation}
and zero components of CS and Maxwell field potentials

\begin{equation}
a_{0}\left( x\right) =\chi \gamma _{1}\left( C_{2}e^{x\xi
_{1}}-C_{1}e^{-x\xi _{1}}\right) +\chi \gamma _{2}\left( C_{4}e^{x\xi
_{2}}-C_{3}e^{-x\xi _{2}}\right) +C_{6}  \tag{24}
\end{equation}

\begin{equation}
A_{0}\left( x\right) =\frac{1}{\xi _{1}}\left( C_{1}e^{-x\xi
_{1}}-C_{2}e^{x\xi _{1}}\right) +\frac{1}{\xi _{2}}\left( C_{3}e^{-x\xi
_{2}}-C_{4}e^{x\xi _{2}}\right) +C_{7}  \tag{25}
\end{equation}

In Eqs. (24) and (25) two new independent coefficients, $C_{6}$ and $C_{7}$,
appear. These coefficients, which give the asymptotic configurations of the
potentials $a_{0}$ and $A_{0}$ respectively, are related through Eq. (5) to $%
C_{5}$ as follows

\begin{equation}
e\mathit{\Pi }_{\mathit{1}}C_{5}=-\mathit{\Pi }_{\mathit{0}}\left(
C_{6}+eC_{7}\right)  \tag{26}
\end{equation}

Eq. (26) establishes a connection between a linear combination of the
coefficients of the long-range modes of the zero components of the gauge
potentials, $(C_{6}+eC_{7})$, and the coefficient of the long-range mode of
the magnetic field, $C_{5}$. Note that if the induced CS coefficient $%
\mathit{\Pi }_{\mathit{1}}$, or the Debye-screening coefficient $\mathit{\Pi 
}_{\mathit{0}}$ were zero, there would be no link between $C_{5}$ and $%
(C_{6}+eC_{7})$. This relation between the long-range modes of $B$, $A_{0}$
and $a_{0}$ can be interpreted as a sort of Aharonov-Bohm effect, which
occurs in this system at finite temperature\cite{Our}$.$ At $T=0$, we have $%
\mathit{\Pi }_{\mathit{0}}=0$, and this effect disappears.

Then, after using the boundary conditions (17) and the constraint equation
(26), it is found that the leading contribution to the free-energy density $%
\mathit{f}=\frac{\mathcal{F}}{\mathcal{A}}$ ,\ ($\mathcal{A}=LL^{\prime }$
being the sample area) in the sample's length limit $(L\rightarrow \infty $, 
$L^{\prime }\rightarrow \infty )$ is given as a function of $A$ and $C_{1}$
by

\begin{equation}
f=\frac{1}{2}\left[
X_{1}A^{2}+X_{2}C_{1}^{2}+X_{3}AC_{1}+X_{4}A+X_{5}C_{1}+X_{6}\right] 
\tag{27}
\end{equation}
The coefficients $X_{i}$ are expressed in terms of the polarization operator
coefficients as

\[
X_{1}=g\overline{\gamma }_{2}^{2}+\mathcal{G},\qquad X_{2}=g\gamma _{1}^{2}+%
\mathcal{G},\qquad X_{3}=-2g\gamma _{1}\overline{\gamma }_{2},\qquad
X_{4}=-2g\overline{B}\overline{\gamma }_{2}, 
\]

\begin{equation}
X_{5}=2g\overline{B}\gamma _{1},\qquad X_{6}=g\overline{B}^{2}  \tag{28}
\end{equation}

\begin{equation}
g=1+\frac{e^{2}\mathit{\Pi }_{\mathit{1}}\,^{2}}{\mathit{\Pi }_{\mathit{0}}}%
+e^{2}\mathit{\Pi }_{\,\mathit{2}}  \tag{29}
\end{equation}

\begin{eqnarray}
\mathcal{G} &=&\frac{1}{2}\left( 1+\overline{\gamma }_{2}^{2}-\frac{N}{\pi }%
\chi ^{2}\overline{\xi }_{2}\overline{\gamma }_{2}\right) -\left( \frac{%
\mathit{\Pi }_{\mathit{0}}}{2\overline{\xi }_{2}^{2}}+\frac{\mathit{\Pi }_{%
\mathit{0}}\,^{\prime }}{2}\right) \left( e+\chi \overline{\xi }_{2}%
\overline{\gamma }_{2}\right) ^{2}  \nonumber \\
&&-\frac{\mathit{\Pi }_{\mathit{1}}}{\overline{\xi }_{2}}\left( \chi 
\overline{\xi }_{2}-e\overline{\gamma }_{2}\right) \left( e+\chi \overline{%
\xi }_{2}\overline{\gamma }_{2}\right) +\frac{\mathit{\Pi }_{\,\mathit{2}}}{2%
}\left( \chi \overline{\xi }_{2}-e\overline{\gamma }_{2}\right) ^{2} 
\tag{30}
\end{eqnarray}

The values of $A$ and $C_{1}$ are found by minimizing the corresponding
free-energy density

\begin{equation}
\frac{\delta \mathit{f}}{\delta A}=\frac{1}{2}\left(
2X_{1}A+X_{3}C_{1}+X_{4}\right) =0  \tag{31}
\end{equation}

\begin{equation}
\frac{\delta \mathit{f}}{\delta C_{1}}=\frac{1}{2}\left(
2X_{2}C_{1}+X_{3}A+X_{5}\right) =0,  \tag{32}
\end{equation}
to be

\begin{equation}
A=\frac{\overline{\gamma }_{2}}{\gamma _{1}^{2}+\overline{\gamma }_{2}^{2}}%
\overline{B}  \tag{33}
\end{equation}

\begin{equation}
C_{1}=-\frac{g\gamma _{1}^{3}}{\left( g\gamma _{1}^{2}+\mathcal{G}\right)
\left( \gamma _{1}^{2}+\overline{\gamma }_{2}^{2}\right) }\overline{B} 
\tag{34}
\end{equation}

Taking into account the boundary conditions (17) we obtain for the
long-range mode of the magnetic field

\begin{equation}
C_{5}=\gamma _{1}C_{1}-\overline{\gamma }_{2}A+\overline{B}=\frac{\gamma
_{1}^{2}\mathcal{G}}{\left( g\gamma _{1}^{2}+\mathcal{G}\right) \left(
\gamma _{1}^{2}+\overline{\gamma }_{2}^{2}\right) }\overline{B}  \tag{35}
\end{equation}

From Eqs. (33)-(35) it is clear that at $T\gg \omega _{c}$ the unknown
coefficients $A$, $C_{1}$ and $C_{5}$ are all different from zero. At the
densities under consideration, $n_{e}\ll m^{2}$, the estimated values of the
coefficients $A$, $C_{1}$ and $C_{5}$ in the high-temperature approximation $%
\left( T\gg \omega _{c}\right) $ are

\begin{equation}
A\approx 10^{3}\overline{B},\qquad C_{1}\approx -10^{-11}\overline{B},\qquad
C_{5}\approx 10^{-4}\overline{B}  \tag{36}
\end{equation}

These results imply that in the high-temperature phase there exist an
inhomogeneous penetration ($A\neq 0$) and a significantly smaller
homogeneous magnetic penetration ($C_{5}\neq 0$). Note the essentially
different behavior of this system in the low-temperature \cite{Our} and in
the high temperature regions. In other words, the different from zero values
of the coefficients $A$ and $C_{5}$ mean that the superconducting state
existing at $T\ll \omega _{c}$ is broken in the ($T\gg \omega _{c}$)-phase.

Eq. (36), together with Eqs. (18) and (19) yield the following leading
contribution to the magnetic and electric fields respectively

\begin{equation}
B\left( x\right) =\overline{B}\cos \left( \frac{2\pi }{\lambda }x\right) 
\tag{37}
\end{equation}

\begin{equation}
E\left( x\right) =E_{0}\left( T\right) \sin \left( \frac{2\pi }{\lambda }%
x\right)  \tag{38}
\end{equation}

where

\begin{equation}
E_{0}\left( T\right) =\frac{12\sqrt{2}m}{\overline{\xi }_{2}}\left( \tanh 
\frac{\beta \mu }{2}+1\right) ^{-1}\overline{B}  \tag{39}
\end{equation}

\begin{equation}
\lambda =\frac{2\pi }{\overline{\xi }_{2}}  \tag{40}
\end{equation}

In writing Eq. (37) we took into account that the exponentially decaying
component of the magnetic field (the one associated with the coefficient $%
\gamma _{1}C_{1}$ in the general solution (18)), as well as the uniform one,
Eq. (35), are negligible in comparison with the inhomogeneous component
associated with the coefficient $A$.

Hence, the applied magnetic field penetrates the charged anyon fluid at $%
T\gg \omega _{c}$ with a magnitude that essentially changes sinusoidally
with $x$ and is characterized by a wavelength $\lambda $, which depends on
temperature through the length scale $\overline{\xi }_{2}$ (Eq. (40)). At $%
T\gtrsim \omega _{c}$, using that \cite{8} $\mu \simeq \dfrac{\pi n_{e}}{m}$%
, one can estimate that $\lambda \simeq 0.4$ $A^{o}$. On the other hand,
taking into account that $\overline{\xi }_{2}$ increases with the
temperature (see Eq. (16)), we have, that the wavelength decreases with $T$,
having a high-temperature leading behavior given by

\begin{equation}
\lambda \approx \frac{\pi }{24}\sqrt{\frac{1}{2mT}}  \tag{41}
\end{equation}

Finally, let us calculate the induced electric charge density of the charged
medium in the high-temperature approximation. Considering Eq. (9) in the
high-temperature limit, we find that the induced electric charge density
presents an inhomogeneous spatial distribution with high-temperature leading
contribution given by

\begin{equation}
eJ_{0}\left( x\right) =24\sqrt{2}m\overline{B}\left[ \tanh \left( \frac{%
\beta \mu }{2}\right) +1\right] ^{-1}\cos \left( \frac{2\pi }{\lambda }%
x\right)  \tag{42}
\end{equation}

Because in the high-temperature regime $\lambda \sim \sqrt{1/T}$, it follows
that the inhomogeneity of the charge density (42) increases with
temperature. This spatial periodicity of the charge density in the
high-temperature phase is what we pointed out above as a sort of Wigner
crystallization.

In the same way, if we calculate the current density (10) in the
high-temperature limit we find

\begin{equation}
eJ_{2}\left( x\right) =-96\pi \sqrt{2}\sqrt{m/\beta }\overline{B}\cosh
\left( \frac{\beta \mu }{2}\right) \left[ \tanh \left( \frac{\beta \mu }{2}%
\right) +1\right] ^{-\frac{1}{2}}\sin \left( \frac{2\pi }{\lambda }x\right) 
\tag{43}
\end{equation}

Obviously, the current density (43) is not a supercurrent confined to the
sample's boundary.

To conclude, we want to stress that the obtained results indicate that in
the charged anyon fluid, when temperatures larger than the energy gap are
reached, a new phase, on which the superconductivity is lost, appears (no
Meissner effect is found in this phase).

We must point out that the absence of Meissner effect at $T\gg \omega _{c}$
is related to the appearing of an imaginary magnetic mass for the
electromagnetic field in this phase. As we have proved (this result will be
publish elsewhere), the existence of an imaginary magnetic mass cannot be
associated with a tachyonic mode in this many-particle system with CS
interactions. The reason is that magnetic masses and rest energies of
electromagnetic field modes are not equivalent within the charged anyon
fluid.

\textbf{Acknowledgments}

We are indebted to A. Cabo for a very useful discussion. This research has
been supported in part by the National Science Foundation under Grant No.
PHY-9722059.

\end{document}